

\input harvmac.tex

\def\NPB#1{Nucl. Phys. B{#1}}
\def\PLB#1{Phys. Lett. B{#1}}
\def\PRD#1{Phys. Rev. D{#1}}
\def\PRL#1{Phys. Rev. Lett. {#1}}

\def\hc{ {\rm h.c.} }
\def\ME{ M_{ETC} }
\def\gE{ g_{ETC} }
\def\bb{ {b{\bar b}} }
\def\Zbb{ {Zb{\bar b}} }
\def\sw{ s_\theta }
\def\cw{ c_\theta }
\def\esc{ {e\over\sw\cw} }
\noblackbox
\def\half{{1 \over 2}}
\def\gae{\raise-.5ex\vbox{\hbox{$\; >\;$}\vskip-2.9ex\hbox{$\; \sim\;$}}}
\def\Boxmark#1#2#3{\global\setbox0=\hbox{\lower#1em \vbox{\hrule height#2em
     \hbox{\vrule width#2em height#3em \kern#3em \vrule width#2em}%
     \hrule height#2em}}%
     \dimen0=#2em \advance\dimen0 by#2em \advance\dimen0 by#3em
     \wd0=\dimen0 \ht0=\dimen0 \dp0=0pt
     \mkern1.5mu \box0 \mkern1.5mu }

\Title{\vbox{\baselineskip12pt\hbox{BUHEP-92-12}\hbox{HUTP-92/A017}
	\hbox{hep-ph@xxx/9204214}}}
{\vbox{\centerline{Non-Oblique Effects in the $\Zbb$}
	\vskip8pt\centerline{Vertex from ETC Dynamics}}}

\centerline{R. Sekhar Chivukula$^{a*}$, Stephen B. Selipsky$^{a*}$ and
Elizabeth H. Simmons$^{b*}$}
\footnote{}{$^*$ sekhar@weyl.bu.edu,
 sbs@weyl.bu.edu, simmons@huhepl.harvard.edu}
\vskip .1in
\centerline{$^a$ Boston University}
\centerline{Department of Physics}
\centerline{590 Commonwealth Avenue}
\centerline{Boston, MA 02215}
\vskip .1in
\centerline{$^b$ Lyman Laboratory of Physics}
\centerline{Harvard University}
\centerline{Cambridge, MA 02138}
\vskip .2in
\centerline{\bf ABSTRACT}

Extended technicolor theories generate potentially large corrections to the
$\Zbb$ vertex which can be observed in current experiments at LEP.

\Date{04/92}

There exist no compelling or even consistent theories to explain the origin of
the diverse masses and mixings of the quarks and leptons.  In this regard, the
origin of the large top mass is particularly puzzling.  In technicolor models
\ref\techni{ S. Weinberg, \PRD13 (1976) 974 and \PRD19 (1979) 1277\semi L.
Susskind, \PRD20 (1979) 2619.}, this large top mass is presumably the result
of extended technicolor (ETC) \ref\extc{ S. Dimopoulos and L. Susskind,
\NPB155 (1979) 237\semi E. Eichten and K. Lane, \PLB90 (1980) 125. } dynamics
at relatively low energy scales\foot{This is true so long as there are no
additional light scalar particles coupling to ordinary and techni- fermions
\ref\compositehiggs{ T. Appelquist, T. Takeuchi, M. Einhorn and L.C.R.
Wijewardhana, \PLB220 (1989) 223\semi T. Takeuchi, \PRD40 (1989) 2697\semi
V.A. Miranksy and K. Yamawaki, Mod. Phys. Lett. A4 (1989) 129\semi R.  S.
Chivukula, A. G. Cohen, and K. Lane, \NPB343 (1990) 554.} \ref\simmons{E. H.
Simmons, \NPB312 (1989) 253.}.}.  Since the magnitude of the KM matrix element
$|V_{tb}|$ is very nearly one, $SU(2)_W$ gauge invariance insures that the ETC
dynamics which generates the top mass also couples to the left-handed
component of the bottom.  In this note, we point out that this dynamics
produces potentially large ``non-oblique'' \ref\oblique{ B.W. Lynn, M.E.
Peskin and R.G. Stuart, SLAC-PUB-3725 (1985); in {\it Physics at LEP}, Yellow
Book CERN 86-02, Vol. I p.90. } effects at the $\Zbb$ vertex.  In particular,
if $m_t \gae 100$ GeV and no effect is visible with data currently being
obtained at LEP, theories in which the ETC and weak interactions commute ({\it
i.e.} in which the ETC gauge bosons are $SU(2)_W$ singlets) can be ruled out,
with the same confidence as models with excessive flavor changing neutral
currents.

If the top mass is generated by the exchange of an $SU(2)_W$ neutral ETC gauge
boson, then this boson carries technicolor and couples with strength $\gE$ to
the current
  \eqn\tmasscur{ \xi {\bar\psi^i}_L \gamma^\mu T_L^{iw} +
   {1\over \xi} {\bar t_R} \gamma^\mu U_R^w ,}
where $\psi_L = (t,b)_L$ is the left-handed $tb$ doublet, $T_L = (U,D)_L$
is a left-handed technifermion weak doublet, and $U_R$ is a corresponding
right-handed technifermion weak singlet.  The indices $i$ and $w$ are for
$SU(2)_W$ and technicolor, respectively.  The constant $\xi$ is an ETC
gauge-group dependent Clebsch and is expected to be of order one.  At energies
lower than the mass ($\ME$) of the ETC gauge boson, the effects of its
exchange may be approximated by local four-fermion operators.  In particular,
the top mass arises from an operator coupling the left- and right- handed
pieces of the current \tmasscur\
  \eqn\topff{ -{\gE^2 \over \ME^2}  \left({\bar\psi}_L^i \gamma^\mu
T_L^{iw}\right) \left( {\bar U^w}_R \gamma_\mu t_R \right) + \hc\ .}
This may be Fierzed into a product of technicolor singlet densities
  \eqn\topfierz{ 2 {\gE^2\over\ME^2} \left( {\bar\psi^i}_L
   t_R \right) \left( {\bar U}_R T^i_L \right) + \hc\ .}

In what follows we will assume (for simplicity) that there is only one doublet
of technifermions, that the strong technicolor interactions respect an
$SU(2)_L \times SU(2)_R$ chiral symmetry, and therefore that the technicolor
$F$ constant (analogous to $f_\pi$ in QCD) is $v\approx 250$ GeV.  Using the
rules of naive dimensional analysis \ref\nda{ A. Manohar and H. Georgi,
\NPB{234} (1984) 189. } we find the top quark mass is
  \eqn\topmass{ m_t\ = {\gE^2 \over \ME^2}
   \langle{\bar U}U\rangle\ \approx\ {\gE^2 \over \ME^2} (4\pi v^3)\ .}
Equivalently, the scale of the ETC dynamics responsible for generating
the top mass is
  \eqn\etcmass{ \ME \approx 1.4\ {\rm TeV} \cdot
   \gE \left({100\ {\rm GeV} \over m_t} \right)^\half\ .}
In the absence of fine tuning \compositehiggs\ and as long as $\gE^2 v^2/\ME^2
< 1$ (or, equivalently, when $m_t/(4\pi v)$ is small), the ETC interactions
may be treated as a small perturbation on the technicolor dynamics and our
estimates are self consistent.

These dimensional estimates are typically modified in ``walking technicolor''
models \ref\walking{ B. Holdom, \PLB105 (1985) 301\semi T. Appelquist, D.
Karabali and L.C.R. Wijewardhana, \PRD35 (1987) 2605\semi M. Bando, T.
Morozumi, H. So and K. Yamawaki, \PRL59 (1987) 389\semi V.A. Miransky, Nuovo
Cimento 90A (1985) 149. } where there is an enhancement of operators of the
form \topfierz\ due to a large anomalous dimension of the technifermion mass
operator.  The enhancement is important for the ETC interactions responsible
for light fermion masses (for which $\ME$ must be quite high), but will not be
numerically significant in the case of the top quark because the ETC scale
\etcmass\ associated with the top quark is so low.  Hence, the results in
``walking'' theories are expected to be similar to those presented below.

Consider the four-fermion operator\foot{Ref.~\ref\bigrho{ T. Appelquist, M.J.
 Bowick, E. Cohler and A.I. Hauser, \PRD31 (1985) 1676. } lists possible
four-fermion operators arising from ETC exchange, with emphasis on potentially
dangerous ETC contributions to $\delta\rho$.} arising from the left-handed
part of the current \tmasscur\
\eqn\unfierz{ -\xi^2 {\gE^2\over\ME^2}
\left({\bar \psi^i}_L \gamma^\mu T^{iw}_L \right)
\left({\bar T^{jw}}_L \gamma_\mu \psi^j_L \right)\ .}
This may be Fierzed into the form of a product of technicolor singlet
currents and includes
  \eqn\fourferm{ -{\xi^2\over 2}
   {\gE^2\over \ME^2}\left({\bar\psi}_L\gamma^\mu\tau^a\psi_L \right)
    \left({\bar T}_L \gamma_\mu \tau^a T_L \right)\ ,}
where $\gE$ and $\ME$ are as in eqn.~\topfierz\ and the $\tau^a$ are weak
isospin Pauli matrices.  We will show that this operator can generate sizeable
deviations in the predictions for the $\Zbb$ coupling.  There are also
operators involving products of weak singlet left-handed currents, but
these operators will not affect the $\Zbb$ coupling.

Our analysis of the effects of operator \fourferm\ proceeds along the lines
of ref. \ref\sekharlisa{ L. Randall and R.S. Chivukula, \NPB326 (1989) 1.}.
Adopting an effective chiral Lagrangian description appropriate below the
technicolor chiral symmetry breaking scale, we may replace the technifermion
current by a sigma-model current \ref\howardbook{ H. Georgi, {\it Weak
Interactions and Modern Particle Theory}, (Benjamin-Cummings, Menlo Park,
1984), p.77. }:
  \eqn\interpolate{ \left({\bar T}_L \gamma_\mu \tau^a T_L \right) =
	{v^2 \over 2}Tr\left(\Sigma^\dagger\tau^a iD_\mu\Sigma\right)\ ,}
where $\Sigma = \exp{(2i{\tilde\pi}/v)}$ transforms as
$\Sigma \rightarrow L\Sigma R^\dagger$ under $SU(2)_L \times SU(2)_R$,
and the covariant derivative is
  \eqn\covariant{ \partial_\mu\Sigma
+ i{e\over\sw\sqrt2}\left(W_\mu^+\tau^+ + W_\mu^-\tau^-\right)\Sigma
+ i\esc Z_\mu\left({\tau_3\over 2}\Sigma - \sw^2[Q,\Sigma]\right)
+ ieA_\mu[Q,\Sigma]\ .}
In unitary gauge $\Sigma=1$ and operator \fourferm\ becomes
  \eqn\zslashop{ {\xi^2\over 2} {\gE^2 v^2\over\ME^2}{\bar\psi}_L \left(
 \esc Z\!\!\!\!/ {\tau_3\over 2} +
 {e\over\sw\sqrt2}\left(W\!\!\!\!\!/\ ^+ \tau^+ + W\!\!\!\!\!/\ ^-
  \tau^-\right) \right)\psi_L\ .}
This yields a correction
  \eqn\dgetc{\delta g_L = -{\xi^2 \over 2} {\gE^2 v^2\over\ME^2} \esc(I_3)
	= {\xi^2 \over 4} {m_t\over{4\pi v}} \cdot \esc}
to the tree-level $\Zbb$ couplings $g_L = \esc (I_3-Q\sw^2) =
 \esc(-\half + {1\over 3}\sw^2)$ and $g_R = \esc({1\over 3}\sw^2)$.

The $\Zbb$ width consequently shifts by an amount
  \eqn\dgamma{ {\delta\Gamma\over\Gamma_\bb} \approx { 2{g_L\delta g_L}\over
   {g_L^2+g_R^2}} \approx -3.7\% \cdot \xi^2
   \left({m_t\over 100\ {\rm GeV}}\right)\ .}
Note that the shift is {\it linear} in $m_t$; it is also a non-oblique
correction applying only to the $\Zbb$ width.  As the standard model
prediction for $\Gamma_\bb$ is 378 MeV, we see that \dgamma\ leads to a
reduction of
\eqn\whynot{14\ {\rm MeV} \cdot \xi^2 \left({m_t \over 100\ {\rm GeV}} \right)}
in both $\Gamma_\bb$ and
$\Gamma_{had}$ (the hadronic width).  By way of comparison, the leading
$m_t$-dependent term of the $\Zbb$ vertex correction in the one Higgs standard
model is \ref\smvertex{
 A.A. Akhundov, D.Yu. Bardin and T. Riemann, \NPB276 (1986) 1 \semi
 W. Beenakker and W. Hollik, Z. Phys. C40 (1988) 141\semi
 J. Bernabeu, A. Pich and A. Santamaria, \PLB200 (1988) 569\semi
 B.W. Lynn and R.G. Stuart, \PLB252 (1990) 676. }
  \eqn\dgsm{ \delta g_L = -2 I_3\left({m_t\over 4\pi v}\right)^2 \cdot\esc }
giving an $m_t$-dependance in the $\Zbb$ width of
  \eqn\smdgamma{ {\delta\Gamma\over\Gamma_\bb} \approx
   -0.5\% \left({m_t\over 100\ {\rm GeV}}\right)^2 }
and a correspondingly smaller effect in the hadronic width.  For $m_t \gae
100$ GeV, \smdgamma\ is quite a good estimate of the full one-loop standard
model result.  For a 100 GeV top, \smdgamma\ corrects $\Gamma_\bb$ and
$\Gamma_{had}$ by approximately 2 MeV.

Experiments at LEP currently measure $\Gamma_\bb$ to an accuracy of 5\%
\ref\zbblep{ J. Kroll, XXVIIth Rencontres de Moriond ``Electroweak
Interactions and Unified Theories'', March 1992, to be published} and
$\Gamma_{had}$ to 12 MeV \ref\joeref{The LEP Collaborations, \PLB276 (1992)
247. }, so a shift on the order of \dgamma\ cannot currently be excluded.  The
measurement of $\Gamma_\bb$ should eventually reach 2\% \zbblep, at which
point it will be possible to distinguish between the results \dgamma\ and
\smdgamma.  The similarly large $Wtb$ vertex contribution in eqn.~\zslashop\
is much more difficult to observe without detailed studies of the top quark.

So far, we have assumed that the ETC and weak interactions commute.  In
theories with weak-charged gauge bosons, we can make no definite predictions.
As before, the ETC boson responsible for generating the top mass can
contribute to the $\Zbb$ vertex . For example, the operator \topfierz\ can
arise from the exchange of a weak-doublet ETC gauge boson which couples $T_L$
to $t^c_L$ (the field which is charge-conjugate to $t_R$) and $\psi_L$ to
$U^c_L$.  Such a gauge boson will give rise to the $SU(2)_{L+R}$ triplet
operator $(\bar{U}_R \gamma^\mu U_R) (\bar{\psi}_L \gamma_\mu \psi_L)$.  In
addition, there may be technicolor-neutral weak-triplet ETC bosons
contributing directly to an operator of the form \fourferm. In both of these
cases, we would generically expect effects on the $\Zbb$ coupling to be of the
same order of magnitude as those already described, but the size and sign of
the total shift \dgamma\ will be model-dependent.

It is interesting to note that a correction to the $\Zbb$ vertex linear in
$m_t$ can also occur in models \ref\david{ D.B. Kaplan, \NPB365 (1991) 259. }
where fermion masses arise from mixing of ordinary fermions and
technibaryons.  In this case, the $b_L$ and $b_R$ are partly technibaryons
and, as in QCD, the axial technibaryon coupling is renormalized.  Then the
left- and right-handed couplings receive a correction of the form
  \eqn\davidgl{\delta g_L = (g_L - g_R) \left({g_A - 1 \over 2}\right)
  \sin^2 \alpha\ ,}
and
  \eqn\davidgr{\delta g_R = (g_R - g_L) \left( {g_A - 1 \over 2} \right)
  \sin^2 \beta\ ,}
where $g_A$ is the axial current renormalization, while $\alpha$ and $\beta$
are the mixing angles relating the left- and right-handed components of the
mass eigenstate $b$ field to the corresponding gauge eigenstate $b$ and
technibaryon fields.  In this model, the mass of the top is
  \eqn\davidtop{m_t \approx m_{TB} \sin \alpha \sin \gamma\ ,}
where $m_{TB}$ is the mass of a technibaryon and $\gamma$ is the mixing
angle for the right-handed top.  If $\sin \gamma$ and $\sin \alpha$ are of
the same order of magnitude, $\sin\alpha\approx\sqrt{m_t/m_{TB}}$ and
  \eqn\davidgam{ {\delta\Gamma\over\Gamma}\approx (g_A-1){m_t\over m_{TB}}\ .}
In a QCD-like theory with $g_A\approx 1.25$, and for $m_t = 100$ GeV and
$m_{TB} = 1$ TeV, this results in an effect of order +2.5\%.

\centerline{\bf Acknowledgements}

We thank Andrew Cohen, Howard Georgi, David Kaplan, Joe Kroll, Kenneth Lane,
Jenny Thomas, and Bing Zhou for useful conversations and comments.  R.S.C.
acknowledges the support of an NSF Presidential Young Investigator Award and
of an Alfred P.  Sloan Foundation Fellowship.  S.B.S. and E.H.S. acknowledge
the support of Texas National Research Laboratory Commission SSC Fellowship
Awards.  This work was supported in part under NSF contracts PHY-9057173 and
PHY-8714654 and under DOE contract DE-AC02-89ER40509 and by funds from the
Texas National Research Laboratory Commission under grants RGFY91B6 and
RGFY9106.

\listrefs
\end